# Identifying interdisciplinarity through the disciplinary classification of co-authors of scientific publications

Giovanni Abramo*
*Institute for Systems Analysis and Computer Science-National Research Council of Italy*
*and*
*Laboratory for Studies of Research and Technology Transfer*
*at University of Rome "Tor Vergata" – Italy*
ADDRESS: Dipartimento di Ingegneria dell'Impresa, Università degli Studi di Roma "Tor Vergata", Via del Politecnico 1, 00133 Roma - ITALY
tel. and fax +39 06 72597362, giovanni.abramo@uniroma2.it

Ciriaco Andrea D'Angelo
*Laboratory for Studies of Research and Technology Transfer*
*at University of Rome "Tor Vergata" – Italy*
ADDRESS: Dipartimento di Ingegneria dell'Impresa, Università degli Studi di Roma "Tor Vergata", Via del Politecnico 1, 00133 Roma – ITALY
tel. and fax +39 06 72597362, dangelo@dii.uniroma2.it

Flavia Di Costa
*Laboratory for Studies of Research and Technology Transfer*
*at University of Rome "Tor Vergata" – Italy*
ADDRESS: Dipartimento di Ingegneria dell'Impresa, Università degli Studi di Roma "Tor Vergata", Via del Politecnico 1, 00133 Roma – ITALY
tel. and fax +39 06 72597362, di.costa@dii.uniroma2.it

\* Corresponding author

# Identifying interdisciplinarity through the disciplinary classification of co-authors of scientific publications[1]

**Abstract**

The growing complexity of challenges involved in scientific progress demands ever more frequent application of competencies and knowledge from different scientific fields. The present work analyzes the degree of collaboration among scientists from different disciplines in order to identify the most frequent "combinations of knowledge" in research activity. The methodology adopts an innovative bibliometric approach based on the disciplinary affiliation of publication co-authors. The field of observation includes all publications (167,179) indexed in the Science Citation Index Expanded (SCI-E) for the five years 2004-2008, authored by all scientists in the hard sciences (43,223) at Italian universities (68). The analysis examines 205 research fields grouped in nine disciplines. Identifying the fields with the highest potential of interdisciplinary collaboration is useful to inform research polices at national and regional levels, as well as management strategies at the institutional level.

# 1. Introduction

One of the striking phenomena of current research activity is the constant growth in collaboration among scientists, favored in part by development of communications technologies. This is witnessed by data concerning scientific production, showing a growing trend towards articles realized in collaboration (Wuchty et al., 2007; Archibugi and Coco, 2004; Persson et al., 2004). The phenomenon has been studied from various viewpoints, with different objectives and methods.

One of the approaches involves exploration of the motives that push researchers to undertake collaborations: to aid solutions to complex problems, search for complementary expertise, gain access to facilities, obtain new research funding or improve research performance, as well as for more personal motivations such as to increase prestige and visibility in the researcher's scientific community (Bozeman and Corley, 2004; Melin, 2000; Katz and Martin, 1997). Huutoniemi et al. (2010) actually hold that scientific collaboration almost always responds to very pragmatic needs, such as reduction of costs or sharing of equipment, which would explain the fact that the resulting research seems multidisciplinary, more than inter- or transdisciplinary.

A second approach, geographical and organizational in nature, investigates the degree of cross-country collaborations (Abramo et al. 2011, Hoekman et al., 2009; Ponds et al., 2007; Wagner and Leydesdorff, 2005.) or cross-organization collaborations (Corley et al., 2006; Chompalov et al., 2002), with emphasis on the analysis of public-private collaboration (Abramo et al., 2010; Bjerregaard, 2010; Abramo et al., 2009a; Etzkowitz, 2003).

A third approach explores the relation between degree of collaboration and research performance (Abramo et al., 2009b, Levitt and Thelwall, M., 2008).

A fourth and final perspective, in which the present work is inserted, investigates the interdisciplinary character of research collaboration. In general, the literature classifies research activity that involves experts of different disciplines under three principle categories: multidisciplinary, interdisciplinary and transdisciplinary (OECD, 1998). Stokols et al. (2003) provide a brief and precise distinction: "Multidisciplinarity refers to a process whereby researchers in different disciplines work independently and sequentially, each from his or her own discipline-specific perspective, to address a common problem. Interdisciplinarity is a process in which researchers work jointly, but from each of their respective disciplinary perspectives, to address a common problem. Transdisciplinarity is a process by which researchers work jointly to develop and use a shared conceptual framework that draws together discipline-specific theories, concepts, and methods to address a common problem".

The literature on the theme of interdisciplinarity is still young, but already seems to show several distinct directions. On the taxonomic front, Klein (2008) proposes a coherent reference framework, structured according to seven generic principles, to classify and evaluate multidisciplinary, interdisciplinary and transdisciplinary research. From a strictly methodological point of view, analysis of the literature reveals that the study of "interdisciplinarity" (understood in the broadest sense, of collaboration among researchers in different disciplines) has been conducted both through field research based on surveys (Sanz et al., 2001; Palmer, 1999, Qin et al., 1997), and with quantitative desk measures obtained through bibliometric approaches and social network analysis (Schummer, 2004).

Bibliometric approaches, such as adopted in the current work, take articles as the subject of study and measure interdisciplinarity in terms of the co-occurrence of discipline-specific items, such as keywords, classification headings, authors, or citations. The most diffuse of these approaches is certainly citation analysis, which can be employed at various levels of analysis: not only individual papers, but also researchers, research teams, organizations, disciplines and nations. A rich literature has flourished in recent years, involving the study of bibliometric networks through mapping and clustering techniques (Waltman et al. 2010; Leydesdorff and Rafols 2009; Noyons and Calero-Medina, 2009; Boyak et al. 2005). Parallel to this there has been development of ever more sophisticated software for elaboration and visualization of such networks (Van Eck and Waltman, 2010).

In bibliometric studies specifically on interdisciplinarity, Rafols and Meyer (2010) hold that the percentage of citations made to other papers outside of the discipline of the paper is the most appropriate base indicator to study the interdisciplinarity of the paper itself. Rinia et al. (2002), in their evaluation of physics research programs in the Netherlands held in 1996, proposed a measure of interdisciplinarity of programs defined as the extent to which articles originating from such programs are published in journals attributed to other disciplines than those belonging to the main discipline. The degree of interdisciplinarity is defined as the percentage of non-main discipline papers. These two works present a common methodological trait that is very important: the assumption that a certain paper inherits the subject category or categories associated with the publishing journal, which in fact in many cases result as being multiple or too general (i.e. Engineering, Mathematics, Medicine, etc.). This assumption is in fact present in various other works on interdisciplinarity (Levitt and Thelwall, 2008; Porter et al. 2008; Adams et al., 2007; Morillo et al., 2003; Morillo et al., 2001): in all of these contributions, interdisciplinarity is a characteristic that is observable and measurable by referring to a standard of classification for the journals observed in the various international bibliometric repertories, such as Thomson Reuters' Web of Science or Elsevier's Scopus. Also Porter and Rafols (2009), who introduce a new index of interdisciplinarity, named the integration score, aimed at describing not only the number of disciplines cited by a paper or their degree of concentration but also their "distance" associate each article in a journal to the subject category of the journal.

The literature offers scarce works that address the problem of bibliometric identification of interdisciplinarity from another perspective: that of co-authorship. Although there would be some differences among different nations, disciplinary groups, scientific communities or single organizations, we can assume that co-authorship is certainly a reliable indication of the contribution of certain scientists to the success of a specific research project. In particular, the discipline to which an author belongs can be envisaged as his/her disciplinary knowledge contribution to that project. Citing Schummer (2004): "co-author analysis measures interdisciplinarity in terms of successful research interaction between disciplines". The problem of interdisciplinary recognition shifts from the semantic analysis of an article or the scientific classification of the cited papers to the identification of its authors' specializations. In his investigations of patterns and degrees of interdisciplinarity in nano-science research, Schummer applied this approach to a dataset of 600 papers published in "nano" journals in the 2002-2003 biennium. His study was based on observation of the disciplinary affiliation of the co-authors of such papers, obtained by their departmental affiliation showed in the title page of each paper or provided by an internet research. By admission of the author, this was a "tedious work", clearly possible only for such a restricted set of papers[2]. In addition to its limited scope (600 papers only), the study suffers from the use of departmental affiliation as a proxy for disciplinary affiliation, which is of questionable validity: in fact departments are organizational units that may embed members of different fields.

The work proposed here overcomes this limitation, notably enlarging the field of observation to all scientific works (173,134) indexed in the Science Citation Index of Thomson Reuters and authored by all Italian university researchers in the hard sciences (43,223) over the five years 2004–2008. The Italian case results as particularly adapted to the bibliometric study of interdisciplinarity based on co-authorship: in the Italian academic system set up by the Ministry of Education, University and Research (MIUR), in what seems practically a unique situation, each academic scientist classifies himself or herself in one and only one scientific field, named a scientific disciplinary sector (SDS, 370 in all)[3]. SDSs are grouped in university disciplinary areas (UDAs, 14 in all). The SDS classification of researchers reflects their educational background, expertise and their prevalent field of research, which does not mean that his or her research is necessarily always confined within his or her SDS (e.g. statisticians may carry out research in medicine, physics, social sciences, etc.). Thus, once the true identities of a publication's co-authors are disambiguated, each publication can be associated with the SDSs (and UDAs) of its authors. The analysis of such associations permits a very robust study of interdisciplinarity, but does not permit discrimination between multidisciplinarity, interdisciplinarity and transdisciplinarity. In fact, co-authorship demonstrates some form of collaboration but does not specify the modality in any way. Thus, from this point on we use the term "interdisciplinarity" to refer to the diversity of the SDSs (or UDAs) associated with the authors of a publication.

The aim of the paper is, through recognition of the disciplinary affiliation of the article co-authors, to achieve identification of the degree of collaboration among disciplines (UDAs) and among fields (SDSs); in the case of identifying the fields this will include distinguishing collaboration between fields of the same discipline and of different disciplines. In particular, the authors intend to identify, beginning from the publications realized by Italian university scientists, the SDSs and thus the UDAs that join with greater frequency in authorship of research works. In this regard, the results of the study provide notable amplification of the scarce material available in the literature. The only information found by the authors is provided by van Rijnsoever and Hessels (2010) who, through a survey of research staff at Utrecht University faculties of science, geosciences and a biomedical cluster, attempted to identify the optimal conditions for interdisciplinary research and the characteristics of researchers associated with disciplinary and interdisciplinary research collaborations. Their study showed that disciplinary collaborations occur more frequently in basic disciplines (relatively autonomous, with the primary objective of advancing fundamental knowledge, i.e. physics and chemistry) while interdisciplinary occurs more in strategic disciplines

---

[2] No existing database provides the affiliation address of every author.

[3] In Italy, MIUR recognizes a total of 95 universities, with the authority to issue legally-recognized degrees. All personnel enter the university system through national public examinations, and career advancement also requires such public examinations. Examinations are given per field (SDS). Members of the examination committee all belong to the same SDS. Candidates need to choose the SDS in which to compete, and show their competence in that specific SDS (through their research outputs). The SDS classification is set up by a very large committee of university professors (named CUN) representing all disciplines in science. The Ministry intervention is limited to enact what CUN proposes.

(more strongly connected with practical applications, i.e. medicine).

For decision makers, at the national, regional and also organizational level, it is exceptionally useful to know which research fields show natural synergies and disciplinary contiguity, or could give root to phenomena of scientific convergence with new disciplines, thanks to a common theoretical or applicative basis. Since communication is an essential part of the research process, and research organization performance is highly dependent on good communication among staff (Allen and Fustfeld, 1975), analyses of collaborations can also be useful for re-planning interior spaces and placement of personnel within a single organization, with the objective of favoring inter-field and inter-discipline collaboration, and thus take advantage of potential economies of scope in research activities (Tassey, 1991).

The next section of this work presents the field of observation and the methodology adopted. Section 3 shows the analyses carried out, while the final section summarizes the main results and discusses their implications.

**2. Data and methodology**

Data used in this work are extracted from the Italian Observatory on Public Research (ORP), a bibliometric database derived from the raw data of the Science Citation Index Expanded. The ORP censuses the scientific production of all researchers employed in Italian public research institutions, starting from 2001. Beginning from this database, we extracted the hard sciences publications authored by Italian universities in the period 2004–2008, amounting to a total of 167,179. When the field of investigation is limited to the hard sciences the literature certainly gives ample justification for the choice of considering scientific publications as a reliable proxy of overall research output (Moed et al., 2004). Furthermore, co-authored publications remain as one of the most tangible and best documented indicators of collaboration in the hard sciences, even though many bibliometricians have cautioned that co-authorship-based indicators should be handled with care as a source of evidence for true scientific collaboration (Lundberg et al., 2006; Laudel 2002; Katz and Martin 1997; Melin and Persson 1996). As Katz and Martin (1997) stated, some forms of collaboration do not generate co-authored articles and some co-authored articles do not reflect actual collaboration. However, for the current study, the ample field of observation permits a certain level of confidence in the findings.

Using an MIUR database[4] concerning all Italian university research staff, and with a complex algorithm developed by D'Angelo et al. (2010) for reconciliation of addresses and disambiguation of the real identity of the authors, it is possible to automatically attribute each publication to the responsible university scientists. In large scale studies the percentage of homonyms is notable, creating a critical problem for disambiguation of the authorship within acceptable margins of error. This is why bibliometric studies are generally carried out at aggregate levels of analysis, such as at the university level, and other analyses conducted at the level of single scientists or research groups are generally limited to a maximum of a few organizations or scientific disciplines, where it is possible to disambiguate manually. To the authors' knowledge the ORP is currently the largest national bibliometric database of disambiguated authorships.

---

[4] http://cercauniversita.cineca.it/php5/docenti/cerca.php. Last accessed Oct. 26, 2011.

As noted, the organization of Italian university personnel provides that each scientist must belong to a specific SDS. Each SDS in turn belongs to a UDA: the hard sciences consist of 9 UDAs (Mathematics and computer sciences, Physics, Chemistry, Earth sciences, Biology, Medicine, Agricultural and veterinary sciences, Civil engineering and architecture, Industrial and information engineering) and 205 SDSs. A list of all SDSs and UDAs, with their acronyms, is shown in Annex 1. Using the disambiguation algorithm, the true authors were identified for all publications. Since each author belongs to a single SDS, it was then possible to associate each publication with the SDSs of the publishing authors. From this publication-SDS link, it was then possible to carry out the count of the combinations of SDSs that occur with greater frequency. Since the focus of the work is on collaboration, the counts refer to the pairs of SDSs relating to the co-authors of each publication. Because co-authors affiliated with foreign institutions are not classified according the SDS system, we were not able to consider foreign collaborations.

We present an example to illustrate the procedure followed to measure the degree of interdisciplinarity in research collaboration. Through application of the disambiguation algorithm to $P_i$ publications, we identify the relative authors $Aut_{ij}$ (i indicates the publication, j indicates the sequence of associated authors). Suppose we have three publications, one with two authors and two with three authors:

$Pub_1$-> $Aut_{11}$; $Aut_{12}$; $Aut_{13}$
$Pub_2$-> $Aut_{21}$; $Aut_{22}$
$Pub_3$-> $Aut_{31}$; $Aut_{32}$; $Aut_{33}$

Since each author is unequivocally associated with a single SDS, each publication will be associated with the corresponding $SDS_{ij}$:

$Pub_1$-> $SDS_{11}$; $SDS_{12}$; $SDS_{13}$
$Pub_2$-> $SDS_{21}$; $SDS_{22}$
$Pub_3$-> $SDS_{31}$; $SDS_{32}$; $SDS_{33}$

We suppose, for this example, that the SDSs are the following: $SDS_{11}$=CHIM/01; $SDS_{12}$=CHIM/02; $SDS_{13}$=CHIM/01; $SDS_{21}$=CHIM/01; $SDS_{22}$=CHIM/06; $SDS_{31}$=CHIM/01; $SDS_{32}$=CHIM/02; $SDS_{33}$=CHIM/06. Thus we can write:

$Pub_1$-> CHIM/01; CHIM/02; CHIM/01
$Pub_2$-> CHIM/01; CHIM/06
$Pub_3$-> CHIM/01; CHIM/02; CHIM/06

At this point we count all the pairs of SDSs (without double counting in a single publication):

the pair CHIM/01-CHIM/02 occurs in 2 publications ($Pub_1$; $Pub_3$)
the pair CHIM/01-CHIM/06 occurs in 2 publications ($Pub_2$; $Pub_3$)
the pair CHIM/02-CHIM/06 occurs in 1 publication ($Pub_3$)

The general degree of interdisciplinarity of a research field is measured as the ratio of number of publications co-authored by researchers belonging to that field with researchers from other fields, to the number of publications authored by researchers belonging to that field. Instead, the specific degree of interdisciplinarity of a field with another specific research field, is measured as the ratio between the number of publications co-authored by researchers from both fields to the number of publications authored by researchers belonging to the first field. In the above example, the general degree of interdisciplinarity of CHIM/01 would be 100%; while the degree of interdisciplinarity of CHIM/01 with CHIM/02 is 66.6%. A novel element and strong point of this study, compared to previous literature, is the ability to conduct this analysis on a large scale, considering the whole of publications produced by Italian university scientists belonging to the hard sciences SDSs.

## 3 Analysis and results

For the purposes of analyzing the degree of collaboration between scientific fields and to identify the most frequent "combinations of knowledge" occurring in Italian university scientific publications, we carried out an analysis at two levels: the UDA and the SDS.

### 3.1 Analysis of interdisciplinarity at the discipline level (UDA)

From the analysis at the level of UDA (Table 1), we observe that Biology has the greatest percentage of publications (43.9%) in collaboration with other UDAs, equal to almost half its total produced. This figure is a full 15 percentage points higher than the second UDA, which is Chemistry (28.9%). Of the remaining UDAs, five (Physics, Earth sciences, Medicine, Agricultural and veterinary sciences, Civil engineering) show values between 21% and 25%. Values less than 20% occur only in Mathematics and computer sciences (17.0%) and Industrial and information engineering (14.8%).

| UDA | No. Univ. | No. staff | No. WoS publications | of which with other UDAs | of which between different SDS within each UDA |
|---|---|---|---|---|---|
| Mathematics and computer sciences | 65 | 3,285 | 14,038 | 17.0% | 3.3% |
| Physics | 61 | 2,581 | 22,368 | 21.1% | 16.1% |
| Chemistry | 59 | 3,238 | 24,569 | 28.9% | 16.0% |
| Earth sciences | 48 | 1,274 | 4,639 | 20.4% | 14.6% |
| Biology | 66 | 5,160 | 28,021 | 43.9% | 15.6% |
| Medicine | 59 | 11,153 | 50,798 | 24.0% | 29.3% |
| Agricultural and veterinary sciences | 56 | 3,180 | 10,309 | 24.8% | 17.1% |
| Civil engineering | 59 | 3,780 | 4,798 | 24.0% | 4.6% |
| Industrial and information engineering | 68 | 4,863 | 32,086 | 14.8% | 7.3% |

*Table 1: Research staff and publications of Italian universities by discipline (UDA); data 2004-2008*

Data in column 5 provide a measure of interdisciplinarity within each UDA. In Mathematics and computer sciences the percentage of publications co-authored by scientists of different SDSs within this discipline is only 3.3%. On the opposite side, in Medicine the percentage is 29.3%. In all other UDAs, such percentage is intermediate and varying between 4.6% (in Civil engineering) and 17.1% (in Agricultural and veterinary sciences).

The nine UDAs can give rise to 36 combinations of interdisciplinary collaboration. To identify the differing degree of collaboration between UDAs, for every pair we identified all publications by the researchers of each UDA, and those in co-authorship. It was thus possible to calculate both the incidence of the co-authorships in total publications for each of the UDAs, as well as the average incidence for the pair.

The data are presented in Table 2. Column 1 lists all the pairs of UDAs; columns 5 and 6 show the degree of collaboration of each UDA with the others. For each UDA, the shaded values indicate the pairs that show the highest degree of collaboration: column 5 refers to the first UDA of the pair in column 1; column 6 refers to the second. We see, for example, that the researchers of the Mathematics and computer science UDA collaborate most with those in the Industrial and information engineering UDA, while these collaborate most with scientists in Chemistry. The only pair that shows maximum collaboration in both directions is Biology-Medicine, which also has the highest value of average degree of interdisciplinarity. Far back, in terms of values of average incidence, there are the pairs CHIM-BIO (10.4%) and BIO-AGR (8.3%). If we consider the first 10 pairs for average incidence, we observe that the UDAs of Industrial and information engineering and Chemistry enter in four of the pairs, while Biology and Medicine are present three times. The UDAs that are present least are Mathematics and computer sciences and Civil engineering.

| UDA pair | Pub. from 1st UDA (a) | Pub. from 2nd UDA (b) | Joint pub. (c) | Incidence for 1st (d=c/a) | Incidence for 2nd (e=c/b) | Average (d+e)/2 |
|---|---|---|---|---|---|---|
| MAT-FIS | 14,038 | 22,368 | 507 | 3.6% | 2.3% | 2.9% |
| MAT-CHIM | 14,038 | 24,569 | 193 | 1.4% | 0.8% | 1.1% |
| MAT-GEO | 14,038 | 4,639 | 39 | 0.3% | 0.8% | 0.6% |
| MAT-BIO | 14,038 | 28,021 | 272 | 1.9% | 1.0% | 1.5% |
| MAT-MED | 14,038 | 50,798 | 493 | 3.5% | 1.0% | 2.2% |
| MAT-AGR | 14,038 | 10,309 | 37 | 0.3% | 0.4% | 0.3% |
| MAT-ING_CIV | 14,038 | 4,798 | 143 | 1.0% | 3.0% | 2.0% |
| MAT-ING_IND | 14,038 | 32,086 | 981 | 7.0% | 3.1% | 5.0% |
| FIS-CHIM | 22,368 | 24,569 | 1,392 | 6.2% | 5.7% | 5.9% |
| FIS-GEO | 22,368 | 4,639 | 262 | 1.2% | 5.6% | 3.4% |
| FIS-BIO | 22,368 | 28,021 | 811 | 3.6% | 2.9% | 3.3% |
| FIS-MED | 22,368 | 50,798 | 1,034 | 4.6% | 2.0% | 3.3% |
| FIS-AGR | 22,368 | 10,309 | 122 | 0.5% | 1.2% | 0.9% |
| FIS-ING_CIV | 22,368 | 4,798 | 254 | 1.1% | 5.3% | 3.2% |
| FIS-ING_IND | 22,368 | 32,086 | 1,154 | 5.2% | 3.6% | 4.4% |
| CHIM-GEO | 24,569 | 4,639 | 226 | 0.9% | 4.9% | 2.9% |
| CHIM-BIO | 24,569 | 28,021 | 2,717 | 11.1% | 9.7% | 10.4% |
| CHIM-MED | 24,569 | 50,798 | 1,759 | 7.2% | 3.5% | 5.3% |
| CHIM-AGR | 24,569 | 10,309 | 504 | 2.1% | 4.9% | 3.5% |
| CHIM-ING_CIV | 24,569 | 4,798 | 157 | 0.6% | 3.3% | 2.0% |
| CHIM-ING_IND | 24,569 | 32,086 | 1,255 | 5.1% | 3.9% | 4.5% |
| GEO-BIO | 4,639 | 28,021 | 133 | 2.9% | 0.5% | 1.7% |
| GEO-MED | 4,639 | 50,798 | 160 | 3.4% | 0.3% | 1.9% |
| GEO-AGR | 4,639 | 10,309 | 51 | 1.1% | 0.5% | 0.8% |
| GEO-ING_CIV | 4,639 | 4,798 | 80 | 1.7% | 1.7% | 1.7% |
| GEO-ING_IND | 4,639 | 32,086 | 142 | 3.1% | 0.4% | 1.8% |
| BIO-MED | 28,021 | 50,798 | 7,670 | 27.4% | 15.1% | 21.2% |
| BIO-AGR | 28,021 | 10,309 | 1,250 | 4.5% | 12.1% | 8.3% |
| BIO-ING_CIV | 28,021 | 4,798 | 156 | 0.6% | 3.3% | 1.9% |
| BIO-ING_IND | 28,021 | 32,086 | 496 | 1.8% | 1.5% | 1.7% |
| MED-AGR | 50,798 | 10,309 | 765 | 1.5% | 7.4% | 4.5% |
| MED-ING_CIV | 50,798 | 4,798 | 254 | 0.5% | 5.3% | 2.9% |
| MED-ING_IND | 50,798 | 32,086 | 866 | 1.7% | 2.7% | 2.2% |
| AGR-ING_CIV | 10,309 | 4,798 | 53 | 0.5% | 1.1% | 0.8% |
| AGR-ING_IND | 10,309 | 32,086 | 132 | 1.3% | 0.4% | 0.8% |
| ING_CIV-ING_IND | 4,798 | 32,086 | 340 | 7.1% | 1.1% | 4.1% |

*Table 2: Analysis of degree of interdisciplinarity at the discipline (UDA) level; data 2004-2008*

### 3.2 Analysis of interdisciplinarity at the SDS level

We now present the results of analysis of interdisciplinarity at a greater level of detail, among the SDSs. Collaborations between researchers from different SDSs can occur within the same discipline (intra-discipline); or between SDSs from different disciplines (cross-discipline). Particularly in small and medium universities, some SDSs could have a very limited number of researchers and collaboration between SDSs could be induced not only by interdisciplinary potential inherent to the disciplines, but also by the lack of researchers in the same field. For this we carried out correlation analysis between the values of degree of collaboration of the SDSs and number of their researchers. No correlation emerged at the overall level, but significant inverse correlation was found in three UDAs: Chemistry (Spearman correlation index = -0.81), Medicine (-0.37), and Industrial and information engineering (-0.49). In consideration, for these UDAs we limited the interdisciplinary analysis to those SDSs with more than 100 researchers.

As an example, we present the analysis for the SDSs in Chemistry UDA. Here, the CHIM/05 SDS was the only one excluded out of a total 205 analyzed, given its very low number of researchers. After completing the analysis of joint publications at the UDA level, we proceeded to analysis at the greater level of detail represented by the SDSs. From Table 3 we observe that the percentage of publications realized with other SDSs varies between 43.1% (CHIM/06, Organic chemistry) and 75.0% (CHIM/10, Food chemistry). If we distinguish between SDSs belonging to the same UDA and other UDAs, we observe that in eight out 11 cases there is a greater percentage of publications realized with SDS from other UDAs: in such cases, with the exception of CHIM/02, CHIM/03 and CHIM/06, the difference is notable (greater than 18 percentage points). For the remaining three SDSs (CHIM/02, CHIM/04, CHIM/12) there is a more balanced distribution between percentage of intra-UDA and inter-UDA collaborations.

| SDS | CHIM/01 | CHIM/02 | CHIM/03 | CHIM/04 | CHIM/06 | CHIM/07 | CHIM/08 | CHIM/09 | CHIM/10 | CHIM/11 | CHIM/12 |
|---|---|---|---|---|---|---|---|---|---|---|---|
| No. of universities active in the SDS | 43 | 42 | 46 | 23 | 49 | 37 | 29 | 29 | 27 | 14 | 29 |
| No. of researchers in the SDS | 281 | 475 | 620 | 150 | 678 | 191 | 477 | 199 | 65 | 35 | 65 |
| Total no. publications | 2,319 | 5,231 | 6,544 | 1,509 | 5,669 | 2,133 | 2,713 | 1,212 | 580 | 363 | 554 |
| Of which with other SDS (%) | 49.3 | 44.4 | 44.0 | 50.3 | 43.1 | 63.7 | 60.1 | 58.7 | 75.0 | 61.7 | 64.1 |
| Of which with SDSs in the same UDA (%) | 25.3 | 21.7 | 21.0 | 27.2 | 18.5 | 22.8 | 15.6 | 18.1 | 26.2 | 11.3 | 35.4 |
| Of which with SDSs in other UDAs (%) | 24.0 | 22.7 | 23.1 | 23.1 | 24.7 | 40.9 | 44.5 | 40.7 | 48.8 | 50.4 | 28.7 |
| No. of SDSs with which collaborates | 13 | 12 | 12 | 13 | 13 | 14 | 16 | 26 | 24 | 22 | 16 |
| No. of SDSs with which realizes more than 10% of its total production | 0 | 1 | 1 | 2 | 0 | 2 | 1 | 1 | 2 | 1 | 2 |
| No. of UDAs with which collaborates | 3 | 3 | 4 | 3 | 3 | 3 | 2 | 4 | 5 | 5 | 5 |
| No. of UDAs with which realizes more than 10% of its total production | 0 | 0 | 0 | 0 | 0 | 1 | 1 | 0 | 0 | 1 | 0 |

*Table 3: Analysis of degree of interdisciplinarity for SDSs in Chemistry UDA; data 2004-2008*

Another detailed analysis was conducted to identify the SDS pairs with the greatest collaboration (Table 4). As an example we present the examination of SDS CHIM/01 (Analytical chemistry). The data were ordered by percentage value of cross-SDS publications out of total publications by CHIM/01 researchers (column 5). For reasons of space we present only the first 20 pairs: the percentages vary from 9.4% for pair CHIM/01_CHIM/03 to 0.6% for CHIM/01_AGR/16. We see that only four pairings produce more than 100 publications: CHIM/01_CHIM/03 (General and inorganic chemistry), CHIM/01_CHIM/12 (Environmental chemistry and chemistry for cultural heritage), CHIM/01_CHIM/06 (Organic chemistry) and CHIM/01_CHIM/02 (Physical chemistry), and that all these involve SDSs from the same UDA. CHIM/01_CHIM/03 shows a number of publications much greater than the others, with almost double the value of the next ranked pair, CHIM/01_CHIM/12. In the first 20 positions (percentage of joint publications relative to total publications by CHIM/01), half of the SDSs represented (10/20) are external to Chemistry: these are BIO/10, FIS/01, AGR/15, BIO/14, FIS/07, SECS-P/13, ING-IND/22, MED/07, GEO/06, and AGR/16 which belong to seven different UDAs.

| SDS pair | Pub. by 1st SDS (a) | Pub. by 2nd SDS (b) | Joint pub. (c) | Incidence for 1st (d=c/a) | Incidence for 2nd (e=c/b) | Average (d+e)/2 |
|---|---|---|---|---|---|---|
| CHIM/01_CHIM/03 | 2,319 | 6,544 | 218 | 9.4% | 3.3% | 6.4% |
| CHIM/01_CHIM/12 | 2,319 | 554 | 126 | 5.4% | 22.7% | 14.1% |
| CHIM/01_CHIM/06 | 2,319 | 5,669 | 115 | 5.0% | 2.0% | 3.5% |
| CHIM/01_CHIM/02 | 2,319 | 5,231 | 111 | 4.8% | 2.1% | 3.5% |
| CHIM/01_CHIM/10 | 2,319 | 580 | 99 | 4.3% | 17.1% | 10.7% |
| CHIM/01_CHIM/08 | 2,319 | 2,713 | 65 | 2.8% | 2.4% | 2.6% |
| CHIM/01_BIO/10 | 2,319 | 6,161 | 60 | 2.6% | 1.0% | 1.8% |
| CHIM/01_CHIM/09 | 2,319 | 1,212 | 46 | 2.0% | 3.8% | 2.9% |
| CHIM/01_CHIM/07 | 2,319 | 2,133 | 35 | 1.5% | 1.6% | 1.6% |
| CHIM/01_FIS/01 | 2,319 | 8,967 | 31 | 1.3% | 0.3% | 0.8% |
| CHIM/01_AGR/15 | 2,319 | 969 | 31 | 1.3% | 3.2% | 2.3% |
| CHIM/01_BIO/14 | 2,319 | 5,219 | 31 | 1.3% | 0.6% | 1.0% |
| CHIM/01_FIS/07 | 2,319 | 2,671 | 29 | 1.3% | 1.1% | 1.2% |
| CHIM/01_CHIM/04 | 2,319 | 1,509 | 23 | 1.0% | 1.5% | 1.3% |
| CHIM/01_CHIM/11 | 2,319 | 363 | 17 | 0.7% | 4.7% | 2.7% |
| CHIM/01_SECS-P/13* | 2,319 | 139 | 16 | 0.7% | 11.5% | 6.1% |
| CHIM/01_ING-IND/22 | 2,319 | 1,931 | 16 | 0.7% | 0.8% | 0.8% |
| CHIM/01_MED/07 | 2,319 | 2,092 | 14 | 0.6% | 0.7% | 0.6% |
| CHIM/01_GEO/06 | 2,319 | 745 | 13 | 0.6% | 1.7% | 1.2% |
| CHIM/01_AGR/16 | 2,319 | 904 | 13 | 0.6% | 1.4% | 1.0% |

*Table 4: Scope of interdisciplinarity for SDS CHIM/01. First 20 pairings for co-authored publications; data 2004-2008.*
* SDS named SECS-P/13 is "Commodity science" which belongs to social sciences.

After omitting, for reasons of scarce representativity, all SDS pairs that show less than 1% values of incidence of joint publications, Table 5 presents, for each UDA, the SDS that shows the maximum percentage of combined intra- and inter-disciplinary collaborative publications. Values vary from 46.4% incidence for MAT/04 (Complementary Mathematics) to 84.7% for MED/05 (Clinical pathology). Apart from MAT/04, all values are greater than 50% and in two cases out of 9 they actually exceed 70% (MED/05 and BIO/12-Clinical biochemistry and biology).

Further, in five cases out of nine for these SDSs, the percentage of publications realized with SDSs of their own UDA is less than that for publications realized with SDSs in other UDAs: this is seen for Mathematics and computer sciences; Physics; Chemistry; Biology and Civil engineering. The most striking observation here is the extremely narrow focus of collaborations involving Mathematics and computer sciences: the number of collaborating SDSs and thus other UDAs involved is truly limited, particularly if compared to UDAs such as Biology or Medicine. In general, the count of partner SDSs and UDAs clearly diminishes if we consider only a percentage of total production over 10%: in this case, we actually observe that for a full six out of nine UDAs, there were no SDSs involved from outside of the same discipline, demonstrating that the most frequent collaborations remain confined to the same discipline (UDA).

| UDA | Mathematics and computer sciences | Physics | Chemistry | Earth sciences | Biology | Medicine | Agricultural and veterinary sciences | Civil engineering | Industrial and inform. engineering |
|---|---|---|---|---|---|---|---|---|---|
| SDS | MAT/04 | FIS/07 | CHIM/07 | GEO/09 | BIO/12 | MED/05 | VET/08 | ICAR/06 | ING-IND/25 |
| No. of universities active in the SDS | 34 | 45 | 37 | 28 | 42 | 31 | 14 | 38 | 27 |
| No. of researchers in the SDS | 102 | 329 | 191 | 81 | 152 | 114 | 110 | 107 | 106 |
| Total no. publications | 97 | 2,671 | 2133 | 313 | 1,654 | 968 | 386 | 95 | 722 |
| Of which with other SDSs (%) | 46.4 | 62.3 | 63.7 | 62.6 | 77.6 | 84,7 | 69.4 | 35.8 | 51.9 |
| Of which with SDSs of the same UDA (%) | 12.4 | 20.4 | 22,8 | 35.1 | 18.3 | 48,6 | 49.0 | 7.4 | 27.7 |
| Of which with SDSs of other UDAs (%) | 34.0 | 41.9 | 40.9 | 27.5 | 59.3 | 36.2 | 20.5 | 28.4 | 24.2 |
| No. of SDSs with which collaborates | 2 | 20 | 14 | 13 | 34 | 38 | 20 | 22 | 12 |
| No. of SDSs with which realizes more than 10% of its total production | 0 | 1 | 2 | 2 | 2 | 3 | 2 | 1 | 0 |
| No. of UDAs with which collaborates | 2 | 6 | 3 | 4 | 3 | 4 | 2 | 7 | 4 |
| No of UDAs with which realizes more than 10% of its total production | 0 | 0 | 1 | 0 | 1 | 0 | 0 | 1 | 0 |

*Table 5: SDSs in each UDA that show the highest degree of interdisciplinarity; data 2004-2008*

Table 6 presents, for each UDA, the SDSs that show the maximum percentage of publications realized in collaboration with SDSs of other UDAs. Four of these are the same as the SDSs in Table 5. Concerning the maximum degree of cross-UDA interdisciplinarity, we observe that the values range from 28.5% (ICAR/06) to 59.3% (BIO/12). Further, for eight out of nine UDAs, this value is less than 50%; the exception is BIO/12 (Clinical biochemistry and biology), which is a highly applied scientific field. As we expect, the scientific publications are primarily realized with SDSs belonging to the same discipline, particularly in UDAs such as Mathematics and computer sciences, Industrial and information engineering, and Chemistry. Further, in all cases (nine of nine), for each SDS, the percentage of publications realized with SDSs of their own UDA is less than that of publications realized with SDSs of other UDAs. Finally, again as before, for each SDS, there is a sharp drop in numbers if we count only the partner SDSs and UDAs with which each SDS realizes more than 10% of its production.

To provide information useful to policy-makers, research organization management and scholars, Annex 2 lists all the SDS pairs where the first SDS has a degree of interdisciplinarity higher than 10%. We observe the disciplinary proximity at the basis of many pairings and the low number of cross-UDA pairs. To provide a more exhaustive picture of this last type of interdisciplinarity, Annex 3 shows all pairs with SDSs from different UDAs, where degree of interdisciplinarity exceeds 5%. The list excludes pairs where the body of researchers belonging to the first SDS produced less than 100 publications over the reference period. Given data such as this, application of clustering techniques, taking account of the specific conditions of individual research organizations, would permit optimized restructuring of organizational communication networks.

| UDA | Mathematics and computer sciences | Physics | Chemistry | Earth sciences | Biology | Medicine | Agricultural and veterinary sciences | Civil engineering | Industrial and inform. engineering |
|---|---|---|---|---|---|---|---|---|---|
| SDS | MAT/04 | FIS/07 | CHIM/08 | GEO/06 | BIO/12 | MED/04 | AGR/09 | ICAR/06 | ING-IND/22 |
| No. of universities active in the UDA | 34 | 45 | 29 | 28 | 42 | 47 | 19 | 38 | 39 |
| No. of researchers in the UDA | 102 | 329 | 477 | 113 | 152 | 547 | 100 | 107 | 231 |
| Total no. publications | 97 | 2,671 | 2,713 | 745 | 1,654 | 4,422 | 161 | 95 | 1,931 |
| Of which with other SDSs (%) | 46.4 | 62.3 | 60.1 | 46.3 | 77.6 | 66.5 | 58.4 | 35.8 | 46.1 |
| Of which with SDSs in the same UDA (%) | 12.4 | 20.4 | 15.6 | 12.2 | 18.3 | 29.2 | 18.0 | 7.4 | 8.4 |
| Of which with SDSs in other UDAs (%) | 34.0 | 41.9 | 44.5 | 34.1 | 59.3 | 37.3 | 40.4 | 28.4 | 37.6 |
| No. of SDSs with which collaborates | 2 | 20 | 16 | 17 | 34 | 35 | 21 | 22 | 12 |
| No. of SDSs with which realizes more than 10% of its total production | 0 | 1 | 1 | 1 | 2 | 1 | 1 | 1 | 1 |
| No. of UDAs with which collaborates | 2 | 6 | 2 | 6 | 3 | 3 | 4 | 7 | 2 |
| No. of UDAs with which realizes more than 10% of its total production | 0 | 0 | 1 | 0 | 1 | 0 | 1 | 1 | 1 |

*Table 6: SDSs in each UDA that show the highest degree of cross-UDA interdisciplinarity; data 2004-2008*

## 4. Conclusions

Scientific research activity continually demands greater inputs of competencies and knowledge originating from different fields, to confront the increasing complexity of problems. This work contributes to the ample stream of investigation on such themes of multidisciplinarity and interdisciplinarity. The objective is to analyze the degree of collaboration between different fields and identify the most recurrent "combinations of knowledge" seen in the resulting publications.

The analyses were carried out at varying levels of detail for scientific fields and disciplines, based on publications by Italian university scientists for the period 2004-2008. The analysis of the entire Italian university system was possible thanks to a complex algorithm used to reconcile addresses and disambiguate the identity of authors, which automatically attributes each publication to the responsible university scientists. Given that all Italian university research staff are classified by field, the methodology could then reconstruct the link between publication and authors' scientific fields, thus permitting the identification of all instances of pairs of interdisciplinary collaborations. The methodology proposed, based on the field classification of co-authors, overcomes few limits of mapping and clustering techniques, which do not normalize by the intensity of publications across fields and do not weight the direction of collaboration (i.e. the share of total output of a field produced with the other field of the pair). As compared to analyses based on the same methodology, ours represent a step forward, as the only previous attempt suffered from the use of departmental affiliation as a proxy for disciplinary affiliation, which is of questionable validity.

We found that Biology is the discipline which collaborates more with others, followed by Chemistry. Intradisciplinary collaborations are more frequent among fields in Medicine. Mathematics and computer sciences, and Industrial and information engineering are the disciplines which collaborate least, both at inter- and intra-disciplinary level.

The comparison of our findings with those from other studies is not straightforward because of differences in the methodological approaches and the definition of the disciplinary boundaries. In particular, with regard to studies based on the assignation of articles to the subject categories of the journals, Morillo et al. (2003) warn: "the results should be analyzed with caution since they are highly dependent on the ISI classification scheme, which is not perfect". The present study, in spite of the different methodology to identify interdisciplinary research collaborations, shows few results consistent with others from previous works. The high degree of interdisciplinary collaborations of Biology was already found by Qin et al, (1997). While the outstanding intra-discipline collaborations of Medicine was already shown by Morillo et al. (2003) who, differently from us, found it occurring in Engineering too. Furthermore the thesis by van Rijnsoever and Hessels (2010), who suggest that interdisciplinarity is more frequent in disciplines with practical applications, can be confirmed only for Medicine, but not for Engineering.

The implications of this type of analysis concern both strategic and organizational policy. At the broader, strategic level, it is important that the policy maker know the potential synergies among disciplines and the collaborations that appear more possible, in order to formulate actions that favor them. The organizational implications concern research institutions such as universities, which generally present an ample complement of competencies. The identification of fields with high degree of interdisciplinarity can inform the structuring of organizational communication networks, in order to favor potential economies of scope in research activities.

In a coming work, the authors propose to pursue this particular aspect: to verify the presence of potential economies of scope, through the study of links between performance of universities and the numerosity of the scientific fields with high degrees of interdisciplinarity.

**Annex 1 – SDS list**

| Code | Title | UDA |
|---|---|---|
| MAT/01 | Mathematical Logic | Mathematics and computer sciences |
| MAT/02 | Algebra | Mathematics and computer sciences |
| MAT/03 | Geometry | Mathematics and computer sciences |
| MAT/04 | Complementary Mathematics | Mathematics and computer sciences |
| MAT/05 | Mathematical Analysis | Mathematics and computer sciences |
| MAT/06 | Probability and Mathematical Statistics | Mathematics and computer sciences |
| MAT/07 | Mathematical Physics | Mathematics and computer sciences |
| MAT/08 | Numerical Analysis | Mathematics and computer sciences |
| MAT/09 | Operational Research | Mathematics and computer sciences |
| INF/01 | Computer Science | Mathematics and computer sciences |
| FIS/01 | Experimental Physics | Physics |
| FIS/02 | Theoretical Physics, Mathematical Models and Methods | Physics |
| FIS/03 | Physics of matter | Physics |
| FIS/04 | Nuclear and Subnuclear Physics | Physics |
| FIS/05 | Astronomy and Astrophysics | Physics |
| FIS/06 | Physics for Earth and Atmospheric Sciences | Physics |
| FIS/07 | Applied Physics (Cultural Heritage, Environment, Biology and Medicine) | Physics |
| FIS/08 | Didactics and History of Physics | Physics |
| CHIM/01 | Analytical Chemistry | Chemistry |
| CHIM/02 | Physical Chemistry | Chemistry |
| CHIM/03 | General and Inorganic Chemistry | Chemistry |
| CHIM/04 | Industrial Chemistry | Chemistry |
| CHIM/05 | Science and Technology of Polymeric Materials | Chemistry |
| CHIM/06 | Organic Chemistry | Chemistry |
| CHIM/07 | Foundations of Chemistry for Technologies | Chemistry |
| CHIM/08 | Pharmaceutical Chemistry | Chemistry |
| CHIM/09 | Applied Technological Pharmaceutics | Chemistry |
| CHIM/10 | Food Chemistry | Chemistry |
| CHIM/11 | Chemistry and Biotechnology of Fermentations | Chemistry |
| CHIM/12 | Environmental Chemistry and Chemistry for Cultural Heritage | Chemistry |
| GEO/01 | Palaeontology and Palaeoecology | Earth sciences |
| GEO/02 | Stratigraphic and Sedimentological Geology | Earth sciences |
| GEO/03 | Structural Geology | Earth sciences |
| GEO/04 | Physical Geography and Geomorphology | Earth sciences |
| GEO/05 | Applied Geology | Earth sciences |
| GEO/06 | Mineralogy | Earth sciences |
| GEO/07 | Petrology and Petrography | Earth sciences |
| GEO/08 | Geochemistry and Volcanology | Earth sciences |
| GEO/09 | Mineral Geological Resources and Mineralogic and Petrographic Applications for the Environment and Cultural Heritage | Earth sciences |
| GEO/10 | Geophysics of Solid Earth | Earth sciences |
| GEO/11 | Applied Geophysics | Earth sciences |
| GEO/12 | Oceanography and Atmospheric Physics | Earth sciences |
| BIO/01 | General Botanics | Biology |
| BIO/02 | Systematic Botanics | Biology |
| BIO/03 | Environmental and Applied Botanics | Biology |

| Code | Title | UDA |
|------|-------|-----|
| BIO/04 | Vegetal Physiology | Biology |
| BIO/05 | Zoology | Biology |
| BIO/06 | Comparative Anatomy and Citology | Biology |
| BIO/07 | Ecology | Biology |
| BIO/08 | Anthropology | Biology |
| BIO/09 | Physiology | Biology |
| BIO/10 | Biochemistry | Biology |
| BIO/11 | Molecular Biology | Biology |
| BIO/12 | Clinical Biochemistry and Biology | Biology |
| BIO/13 | Applied Biology | Biology |
| BIO/14 | Pharmacology | Biology |
| BIO/15 | Pharmaceutic Biology | Biology |
| BIO/16 | Human Anatomy | Biology |
| BIO/17 | Histology | Biology |
| BIO/18 | Genetics | Biology |
| BIO/19 | General Microbiology | Biology |
| MED/01 | Medical Statistics | Medicine |
| MED/02 | History of Medicine | Medicine |
| MED/03 | Medical Genetics | Medicine |
| MED/04 | General Pathology | Medicine |
| MED/05 | Clinical Pathology | Medicine |
| MED/06 | Medical Oncology | Medicine |
| MED/07 | Microbiology and Clinical Microbiology | Medicine |
| MED/08 | Pathological Anatomy | Medicine |
| MED/09 | Internal Medicine | Medicine |
| MED/10 | Respiratory Diseases | Medicine |
| MED/11 | Cardiovascular Diseases | Medicine |
| MED/12 | Gastroenterology | Medicine |
| MED/13 | Endocrinology | Medicine |
| MED/14 | Nephrology | Medicine |
| MED/15 | Blood Diseases | Medicine |
| MED/16 | Rheumatology | Medicine |
| MED/17 | Infectious Diseases | Medicine |
| MED/18 | General Surgery | Medicine |
| MED/19 | Plastic Surgery | Medicine |
| MED/20 | Pediatric and Infant Surgery | Medicine |
| MED/21 | Thoracic Surgery | Medicine |
| MED/22 | Vascular Surgery | Medicine |
| MED/23 | Cardiac Surgery | Medicine |
| MED/24 | Urology | Medicine |
| MED/25 | Psychiatry | Medicine |
| MED/26 | Neurology | Medicine |
| MED/27 | Neurosurgery | Medicine |
| MED/28 | Odonto-Stomalogical Diseases | Medicine |
| MED/29 | Maxillofacial Surgery | Medicine |
| MED/30 | Eye Diseases | Medicine |
| MED/31 | Otorinolaringology | Medicine |
| MED/32 | Audiology | Medicine |
| MED/33 | Locomotory Diseases | Medicine |
| MED/34 | Physical and Rehabilitation Medicine | Medicine |
| MED/35 | Skin and Venereal Diseases | Medicine |

| Code | Title | UDA |
|---|---|---|
| MED/36 | Diagnostic Imaging and Radiotherapy | Medicine |
| MED/37 | Neuroradiology | Medicine |
| MED/38 | General and Specialised Pediatrics | Medicine |
| MED/39 | Child Neuropsychiatry | Medicine |
| MED/40 | Gynaecology and Obstetrics | Medicine |
| MED/41 | Anaesthesiology | Medicine |
| MED/42 | General and Applied Hygiene | Medicine |
| MED/43 | Legal Medicine | Medicine |
| MED/44 | Occupational Medicine | Medicine |
| MED/45 | General, Clinical and Pediatric Nursing | Medicine |
| MED/46 | Laboratory Medicine Techniques | Medicine |
| MED/47 | Nursing and Midwifery | Medicine |
| MED/48 | Neuropsychiatric and Rehabilitation Nursing | Medicine |
| MED/49 | Applied Dietary Sciences | Medicine |
| MED/50 | Applied Medical Sciences | Medicine |
| AGR/01 | Rural economy and evaluation | Agricultural and veterinary sciences |
| AGR/02 | Agronomy and Herbaceous Cultivation | Agricultural and veterinary sciences |
| AGR/03 | General Arboriculture and Tree Cultivation | Agricultural and veterinary sciences |
| AGR/04 | Horticulture and Floriculture | Agricultural and veterinary sciences |
| AGR/05 | Forestry and Silviculture | Agricultural and veterinary sciences |
| AGR/06 | Wood Technology and Woodland Management | Agricultural and veterinary sciences |
| AGR/07 | Agrarian Genetics | Agricultural and veterinary sciences |
| AGR/08 | Agrarian Hydraulics and Hydraulic Forest Management | Agricultural and veterinary sciences |
| AGR/09 | Agricultural Mechanics | Agricultural and veterinary sciences |
| AGR/10 | Rural Construction and Environmental Land Management | Agricultural and veterinary sciences |
| AGR/11 | General and Applied Entomology | Agricultural and veterinary sciences |
| AGR/12 | Plant Pathology | Agricultural and veterinary sciences |
| AGR/13 | Agricultural Chemistry | Agricultural and veterinary sciences |
| AGR/14 | Pedology | Agricultural and veterinary sciences |
| AGR/15 | Food Sciences | Agricultural and veterinary sciences |
| AGR/16 | Agricultural Microbiology | Agricultural and veterinary sciences |
| AGR/17 | General Techniques for Zoology and Genetic Improvement | Agricultural and veterinary sciences |
| AGR/18 | Animal Nutrition and Feeding | Agricultural and veterinary sciences |
| AGR/19 | Special Techniques for Zoology | Agricultural and veterinary sciences |
| AGR/20 | Animal Husbandry | Agricultural and veterinary sciences |
| VET/01 | Anatomy of Domestic Animals | Agricultural and veterinary sciences |
| VET/02 | Veterinary Physiology | Agricultural and veterinary sciences |
| VET/03 | General Pathology and Veterinary Pathological Anatomy | Agricultural and veterinary sciences |
| VET/04 | Inspection of Food Products of Animal Origin | Agricultural and veterinary sciences |
| VET/05 | Infectious Diseases of Domestic Animals | Agricultural and veterinary sciences |
| VET/06 | Parasitology and Parasitic Animal Diseases | Agricultural and veterinary sciences |
| VET/07 | Veterinary Pharmacology and Toxicology | Agricultural and veterinary sciences |
| VET/08 | Clinical Veterinary Medicine | Agricultural and veterinary sciences |
| VET/09 | Clinical Veterinary Surgery | Agricultural and veterinary sciences |
| VET/10 | Clinical Veterinary Obstetrics and Gynaecology | Agricultural and veterinary sciences |
| ICAR/01 | Hydraulics | Civil engineering |
| ICAR/02 | Maritime Hydraulic Construction and Hydrology | Civil engineering |
| ICAR/03 | Environmental and Health Engineering | Civil engineering |

| Code | Title | UDA |
|---|---|---|
| ICAR/04 | Road, Railway and Airport Construction | Civil engineering |
| ICAR/05 | Transport | Civil engineering |
| ICAR/06 | Topography and Cartography | Civil engineering |
| ICAR/07 | Geotechnics | Civil engineering |
| ICAR/08 | Construction Science | Civil engineering |
| ICAR/09 | Construction Techniques | Civil engineering |
| ICAR/10 | Technical Architecture | Civil engineering |
| ICAR/11 | Building Production | Civil engineering |
| ICAR/12 | Architecture Technology | Civil engineering |
| ICAR/13 | Industrial Design | Civil engineering |
| ICAR/14 | Architectural and Urban Composition | Civil engineering |
| ICAR/15 | Landscape Architecture | Civil engineering |
| ICAR/16 | Interior Architecture and Venue Design | Civil engineering |
| ICAR/17 | Design | Civil engineering |
| ICAR/18 | History of Architecture | Civil engineering |
| ICAR/19 | Restoration | Civil engineering |
| ICAR/20 | Urban Planning | Civil engineering |
| ICAR/21 | Urban Studies | Civil engineering |
| ICAR/22 | Cadastral Surveying | Civil engineering |
| ING-IND/01 | Naval Architecture | Industrial and information engineering |
| ING-IND/02 | Naval and Marine construction and installation | Industrial and information engineering |
| ING-IND/03 | Flight Mechanics | Industrial and information engineering |
| ING-IND/04 | Aerospace construction and installation | Industrial and information engineering |
| ING-IND/05 | Aerospace Systems | Industrial and information engineering |
| ING-IND/06 | Fluid Dynamics | Industrial and information engineering |
| ING-IND/07 | Aerospatial Propulsion | Industrial and information engineering |
| ING-IND/08 | Fluid Machines | Industrial and information engineering |
| ING-IND/09 | Energy and Environmental Systems | Industrial and information engineering |
| ING-IND/10 | Technical Physics | Industrial and information engineering |
| ING-IND/11 | Environmental Technical Physics | Industrial and information engineering |
| ING-IND/12 | Mechanical and Thermal Measuring Systems | Industrial and information engineering |
| ING-IND/13 | Applied Mechanics for Machinery | Industrial and information engineering |
| ING-IND/14 | Mechanical Design and Machine Building | Industrial and information engineering |
| ING-IND/15 | Design and Methods for Industrial Engineering | Industrial and information engineering |
| ING-IND/16 | Production Technologies and Systems | Industrial and information engineering |
| ING-IND/17 | Industrial and Mechanical Plant | Industrial and information engineering |
| ING-IND/18 | Nuclear Reactor Physics | Industrial and information engineering |
| ING-IND/19 | Nuclear Plants | Industrial and information engineering |
| ING-IND/20 | Nuclear Measurement Tools | Industrial and information engineering |
| ING-IND/21 | Metallurgy | Industrial and information engineering |
| ING-IND/22 | Science and Technology of Materials | Industrial and information engineering |
| ING-IND/23 | Applied Physical Chemistry | Industrial and information engineering |
| ING-IND/24 | Principles of Chemical Engineering | Industrial and information engineering |
| ING-IND/25 | Chemical Plants | Industrial and information engineering |
| ING-IND/26 | Theory of Development for Chemical Processes | Industrial and information engineering |
| ING-IND/27 | Industrial and Technological Chemistry | Industrial and information engineering |
| ING-IND/28 | Excavation Engineering and Safety | Industrial and information engineering |
| ING-IND/29 | Raw Materials Engineering | Industrial and information engineering |
| ING-IND/30 | Hydrocarburants and Fluids of the Subsoil | Industrial and information engineering |
| ING-IND/31 | Electrotechnics | Industrial and information engineering |
| ING-IND/32 | Electrical Convertors, Machines and Switches | Industrial and information engineering |

| Code | Title | UDA |
|---|---|---|
| ING-IND/33 | Electrical Energy Systems | Industrial and information engineering |
| ING-IND/34 | Industrial Bioengineering | Industrial and information engineering |
| ING-IND/35 | Engineering and Management | Industrial and information engineering |
| ING-INF/01 | Electronics | Industrial and information engineering |
| ING-INF/02 | Electromagnetic Fields | Industrial and information engineering |
| ING-INF/03 | Telecommunications | Industrial and information engineering |
| ING-INF/04 | Automatics | Industrial and information engineering |
| ING-INF/05 | Data Processing Systems | Industrial and information engineering |
| ING-INF/06 | Electronic and Information Bioengineering | Industrial and information engineering |
| ING-INF/07 | Electric and Electronic Measurement Systems | Industrial and information engineering |

**Annex 2 - SDS pairs with degree of interdisciplinarity greater than 10%**
*Data 2004-2008 for SDSs with at least 100 publications*

| SDS pair | Co-authored publications | Incidence (%) for 1st SDS | Incidence (%) for 2nd SDS |
| --- | --- | --- | --- |
| MED/32_MED/31 | 99 | 48.8 | 12.1 |
| MED/37_MED/26 | 129 | 45.3 | 3.1 |
| MED/34_MED/26 | 56 | 44.1 | 1.3 |
| FIS/04_FIS/01 | 683 | 43.3 | 7.6 |
| MED/29_MED/28 | 174 | 38.4 | 11.4 |
| ING-IND/05_ING-IND/04 | 42 | 35.0 | 17.4 |
| MED/46_MED/09 | 126 | 30.1 | 1.5 |
| AGR/17_AGR/19 | 149 | 29.9 | 17.7 |
| AGR/18_AGR/19 | 133 | 28.5 | 15.8 |
| MED/20_MED/38 | 72 | 27.5 | 2.1 |
| VET/09_VET/03 | 44 | 27.0 | 7.9 |
| MED/05_MED/04 | 254 | 26.2 | 5.7 |
| FIS/03_FIS/01 | 1707 | 26.1 | 19.0 |
| GEO/09_GEO/06 | 81 | 25.9 | 10.9 |
| MED/16_MED/09 | 257 | 25.0 | 3.0 |
| MED/21_MED/08 | 81 | 24.0 | 2.0 |
| ING-IND/11_ING-IND/10 | 78 | 23.7 | 12.7 |
| BIO/11_BIO/10 | 411 | 23.3 | 6.7 |
| BIO/12_BIO/10 | 383 | 23.2 | 6.2 |
| CHIM/12_CHIM/01 | 126 | 22.7 | 5.4 |
| ING-IND/09_ING-IND/08 | 70 | 22.3 | 13.3 |
| MED/14_MED/09 | 229 | 21.7 | 2.7 |
| MED/13_MED/09 | 586 | 21.6 | 6.8 |
| FIS/07_FIS/01 | 571 | 21.4 | 6.4 |
| GEO/01_GEO/02 | 111 | 21.1 | 19.0 |
| ING-IND/18_ING-IND/19 | 24 | 20.7 | 8.2 |
| MED/49_MED/09 | 63 | 20.1 | 0.7 |
| VET/09_VET/08 | 31 | 19.0 | 8.0 |
| MED/12_MED/09 | 312 | 18.8 | 3.6 |
| MED/46_MED/04 | 78 | 18.7 | 1.8 |
| MED/50_MED/36 | 80 | 18.6 | 3.2 |
| ING-INF/07_ING-INF/01 | 198 | 18.1 | 4.5 |
| BIO/17_BIO/16 | 234 | 17.6 | 10.1 |
| MED/21_MED/18 | 59 | 17.5 | 1.4 |
| MED/37_MED/27 | 49 | 17.2 | 6.3 |
| BIO/15_CHIM/06 | 122 | 17.1 | 2.2 |
| CHIM/10_CHIM/01 | 99 | 17.1 | 4.3 |
| MED/49_BIO/10 | 53 | 16.9 | 0.9 |
| MED/22_MED/18 | 68 | 16.4 | 1.7 |
| BIO/02_BIO/03 | 61 | 16.1 | 12.9 |
| VET/08_VET/03 | 62 | 16.1 | 11.2 |
| MED/12_MED/18 | 263 | 15.8 | 6.4 |
| MED/29_BIO/17 | 70 | 15.5 | 5.3 |
| MED/05_MED/09 | 148 | 15.3 | 1.7 |
| CHIM/07_CHIM/03 | 325 | 15.2 | 5.0 |
| MED/23_MED/11 | 108 | 15.1 | 5.0 |
| MED/39_MED/38 | 88 | 15.0 | 2.5 |

| SDS pair | Co-authored publications | Incidence (%) for 1st SDS | Incidence (%) for 2nd SDS |
| --- | --- | --- | --- |
| ING-IND/22_CHIM/07 | 285 | 14.8 | 13.4 |
| MED/06_MED/08 | 187 | 14.7 | 4.7 |
| MED/22_MED/36 | 60 | 14.5 | 2.4 |
| BIO/12_MED/09 | 236 | 14.3 | 2.7 |
| BIO/19_MED/07 | 89 | 14.2 | 4.3 |
| MED/08_MED/18 | 563 | 14.2 | 13.8 |
| VET/07_BIO/14 | 24 | 14.1 | 0.5 |
| AGR/20_AGR/19 | 34 | 14.0 | 4.0 |
| ING-IND/23_CHIM/07 | 42 | 14.0 | 2.0 |
| ING-IND/27_ING-IND/25 | 72 | 13.9 | 10.0 |
| MED/15_MED/09 | 235 | 13.9 | 2.7 |
| MED/17_MED/07 | 174 | 13.9 | 8.3 |
| CHIM/12_CHIM/02 | 76 | 13.7 | 1.5 |
| AGR/04_AGR/02 | 34 | 13.6 | 6.6 |
| MED/11_MED/09 | 289 | 13.4 | 3.4 |
| MED/39_MED/26 | 78 | 13.3 | 1.9 |
| AGR/20_AGR/18 | 32 | 13.2 | 6.9 |
| GEO/11_GEO/10 | 30 | 13.1 | 5.6 |
| CHIM/02_CHIM/03 | 678 | 13.0 | 10.4 |
| MED/01_MED/09 | 184 | 12.7 | 2.1 |
| CHIM/09_CHIM/08 | 150 | 12.4 | 5.5 |
| MED/35_MED/08 | 143 | 12.4 | 3.6 |
| MED/24_MED/08 | 128 | 12.3 | 3.2 |
| MED/37_MED/36 | 35 | 12.3 | 1.4 |
| MED/06_MED/18 | 156 | 12.2 | 3.8 |
| MED/06_MED/04 | 153 | 12.0 | 3.5 |
| MED/36_MED/18 | 299 | 12.0 | 7.3 |
| MED/46_MED/13 | 50 | 12.0 | 1.8 |
| GEO/07_GEO/08 | 64 | 11.9 | 11.0 |
| MED/50_MED/09 | 51 | 11.9 | 0.6 |
| BIO/17_MED/04 | 154 | 11.6 | 3.5 |
| MED/06_MED/09 | 147 | 11.5 | 1.7 |
| CHIM/04_CHIM/02 | 172 | 11.4 | 3.3 |
| FIS/06_FIS/01 | 38 | 11.4 | 0.4 |
| MED/27_MED/08 | 88 | 11.4 | 2.2 |
| AGR/18_AGR/17 | 53 | 11.3 | 10.6 |
| MED/10_MED/09 | 108 | 11.2 | 1.3 |
| MED/03_MED/38 | 160 | 11.1 | 4.6 |
| MED/27_MED/26 | 86 | 11.1 | 2.0 |
| BIO/15_BIO/14 | 79 | 11.0 | 1.5 |
| MED/34_BIO/09 | 14 | 11.0 | 0.4 |
| MED/49_CHIM/03 | 34 | 10.9 | 0.5 |
| MED/50_MED/28 | 47 | 10.9 | 3.1 |
| GEO/12_FIS/06 | 14 | 10.6 | 4.2 |
| ING-IND/27_CHIM/07 | 55 | 10.6 | 2.6 |
| MED/18_MED/09 | 432 | 10.6 | 5.0 |
| CHIM/11_BIO/10 | 38 | 10.5 | 0.6 |
| GEO/09_GEO/07 | 33 | 10.5 | 6.1 |
| MED/49_BIO/12 | 33 | 10.5 | 2.0 |
| ICAR/01_ICAR/02 | 52 | 10.4 | 8.6 |
| MED/04_MED/09 | 460 | 10.4 | 5.3 |
| CHIM/10_CHIM/06 | 60 | 10.3 | 1.1 |
| MED/15_MED/08 | 174 | 10.3 | 4.4 |
| MED/46_BIO/10 | 43 | 10.3 | 0.7 |
| MED/12_MED/08 | 170 | 10.2 | 4.3 |

| SDS pair | Co-authored publications | Incidence (%) for 1st SDS | Incidence (%) for 2nd SDS |
|---|---|---|---|
| MED/29_MED/08 | 46 | 10.2 | 1.2 |
| BIO/08_BIO/18 | 24 | 10.1 | 2.0 |
| BIO/19_BIO/10 | 63 | 10.1 | 1.0 |
| CHIM/04_CHIM/03 | 152 | 10.1 | 2.3 |
| MAT/01_INF/01 | 17 | 10.1 | 0.4 |
| MED/05_MED/13 | 98 | 10.1 | 3.6 |
| CHIM/10_BIO/14 | 58 | 10.0 | 1.1 |
| VET/10_VET/03 | 22 | 10.0 | 4.0 |

**Annex 3 - SDS pairs with a degree of cross-UDA interdisciplinarity greater than 5%**

*Data 2004-2008 for SDS with at least 100 publications*

| SDS pair | Joint publications | Incidence (%) for 1st SDS | Incidence (%) for 2nd SDS |
|---|---|---|---|
| BIO/15_CHIM/06 | 122 | 17.1% | 2.2% |
| MED/49_BIO/10 | 53 | 16.9% | 0.9% |
| CHIM/08_BIO/14 | 448 | 16.5% | 8.6% |
| MED/29_BIO/17 | 70 | 15.5% | 5.3% |
| ING-IND/22_CHIM/07 | 285 | 14.8% | 13.4% |
| BIO/12_MED/09 | 236 | 14.3% | 2.7% |
| BIO/19_MED/07 | 89 | 14.2% | 4.3% |
| VET/07_BIO/14 | 24 | 14.1% | 0.5% |
| ING-IND/23_CHIM/07 | 42 | 14.0% | 2.0% |
| BIO/17_MED/04 | 154 | 11.6% | 3.5% |
| MED/34_BIO/09 | 14 | 11.0% | 0.4% |
| MED/49_CHIM/03 | 34 | 10.9% | 0.5% |
| GEO/12_FIS/06 | 14 | 10.6% | 4.2% |
| ING-IND/27_CHIM/07 | 55 | 10.6% | 2.6% |
| MED/49_BIO/12 | 33 | 10.5% | 2.0% |
| CHIM/11_BIO/10 | 38 | 10.5% | 0.6% |
| MED/46_BIO/10 | 43 | 10.3% | 0.7% |
| CHIM/10_BIO/14 | 58 | 10.0% | 1.1% |
| CHIM/11_ING-IND/25 | 33 | 9.1% | 4.6% |
| CHIM/08_BIO/10 | 241 | 8.9% | 3.9% |
| ING-IND/24_CHIM/07 | 55 | 8.7% | 2.6% |
| INF/01_ING-INF/05 | 417 | 8.7% | 8.5% |
| BIO/13_MED/04 | 128 | 8.5% | 2.9% |
| CHIM/09_BIO/14 | 102 | 8.4% | 2.0% |
| BIO/17_MED/09 | 111 | 8.4% | 1.3% |
| MED/49_BIO/09 | 26 | 8.3% | 0.7% |
| BIO/16_MED/09 | 192 | 8.3% | 2.2% |
| MED/37_BIO/10 | 22 | 7.7% | 0.4% |
| VET/02_BIO/10 | 32 | 7.5% | 0.5% |
| MED/46_BIO/12 | 31 | 7.4% | 1.9% |
| BIO/16_MED/04 | 172 | 7.4% | 3.9% |
| BIO/17_MED/08 | 98 | 7.4% | 2.5% |
| ING-INF/01_FIS/01 | 325 | 7.3% | 3.6% |
| CHIM/07_FIS/01 | 155 | 7.3% | 1.7% |
| GEO/06_FIS/01 | 54 | 7.2% | 0.6% |
| VET/05_MED/07 | 39 | 7.2% | 1.9% |
| MED/19_BIO/16 | 20 | 7.1% | 0.9% |
| ICAR/03_ING-IND/25 | 23 | 7.1% | 3.2% |
| ING-IND/12_FIS/01 | 17 | 7.1% | 0.2% |
| AGR/20_BIO/06 | 17 | 7.0% | 1.2% |
| MED/03_BIO/13 | 101 | 7.0% | 6.7% |
| BIO/15_CHIM/08 | 50 | 7.0% | 1.8% |
| FIS/07_BIO/10 | 185 | 6.9% | 3.0% |
| BIO/19_CHIM/08 | 43 | 6.9% | 1.6% |
| AGR/01_MED/09 | 13 | 6.8% | 0.2% |
| VET/10_BIO/10 | 15 | 6.8% | 0.2% |
| CHIM/10_BIO/15 | 39 | 6.7% | 5.5% |
| MED/46_BIO/16 | 28 | 6.7% | 1.2% |
| AGR/07_BIO/04 | 28 | 6.6% | 4.9% |
| VET/04_BIO/10 | 15 | 6.6% | 0.2% |
| BIO/16_MED/08 | 152 | 6.5% | 3.8% |
| AGR/12_BIO/10 | 48 | 6.4% | 0.8% |

| SDS pair | Joint publications | Incidence (%) for 1st SDS | Incidence (%) for 2nd SDS |
|---|---|---|---|
| GEO/09_FIS/01 | 20 | 6.4% | 0.2% |
| CHIM/09_BIO/10 | 77 | 6.4% | 1.2% |
| ING-IND/22_FIS/01 | 122 | 6.3% | 1.4% |
| VET/01_MED/17 | 30 | 6.3% | 2.4% |
| MED/44_BIO/14 | 48 | 6.2% | 0.9% |
| MED/27_BIO/10 | 48 | 6.2% | 0.8% |
| MED/05_BIO/10 | 60 | 6.2% | 1.0% |
| BIO/14_MED/09 | 322 | 6.2% | 3.7% |
| BIO/11_MED/04 | 107 | 6.1% | 2.4% |
| CHIM/11_AGR/16 | 22 | 6.1% | 2.4% |
| BIO/12_MED/13 | 100 | 6.0% | 3.7% |
| ING-IND/18_FIS/07 | 7 | 6.0% | 0.3% |
| BIO/13_MED/09 | 90 | 6.0% | 1.0% |
| ING-INF/06_BIO/09 | 67 | 6.0% | 1.9% |
| MED/10_BIO/14 | 56 | 5.8% | 1.1% |
| AGR/20_ING-IND/08 | 14 | 5.8% | 2.7% |
| MED/20_ING-IND/09 | 15 | 5.7% | 4.8% |
| MED/06_BIO/10 | 72 | 5.7% | 1.2% |
| VET/01_MED/09 | 27 | 5.6% | 0.3% |
| BIO/11_MED/09 | 99 | 5.6% | 1.2% |
| AGR/11_BIO/05 | 22 | 5.6% | 1.1% |
| MED/23_BIO/14 | 40 | 5.6% | 0.8% |
| ING-IND/21_CHIM/07 | 30 | 5.6% | 1.4% |
| ICAR/14_ING-IND/32 | 9 | 5.5% | 1.2% |
| MED/49_BIO/14 | 17 | 5.4% | 0.3% |
| ING-IND/12_FIS/03 | 13 | 5.4% | 0.2% |
| MED/43_BIO/14 | 22 | 5.4% | 0.4% |
| MED/03_BIO/10 | 77 | 5.3% | 1.2% |
| ING-IND/23_CHIM/03 | 16 | 5.3% | 0.2% |
| BIO/12_MED/38 | 88 | 5.3% | 2.5% |
| BIO/09_MED/26 | 185 | 5.3% | 4.4% |
| VET/07_BIO/10 | 9 | 5.3% | 0.1% |
| BIO/19_CHIM/06 | 33 | 5.3% | 0.6% |
| MED/06_BIO/14 | 67 | 5.3% | 1.3% |
| GEO/06_CHIM/02 | 39 | 5.2% | 0.7% |
| VET/01_BIO/10 | 25 | 5.2% | 0.4% |
| AGR/07_BIO/10 | 22 | 5.2% | 0.4% |
| ING-IND/27_CHIM/04 | 27 | 5.2% | 1.8% |
| MED/04_BIO/10 | 227 | 5.1% | 3.7% |
| BIO/19_MED/42 | 32 | 5.1% | 2.6% |
| GEO/06_CHIM/03 | 38 | 5.1% | 0.6% |
| ING-IND/23_MAT/07 | 15 | 5.0% | 0.8% |